\documentclass[%
amsmath,amssymb,
reprint,
]{revtex4-1}

\usepackage{graphicx}           
\usepackage{dcolumn}            
\usepackage{bm}                 
\usepackage{color}
\usepackage{tabu}
\usepackage{hyperref}          

\newcommand{\nn}{\nonumber}

\begin{document}

\preprint{}

\title{Implications of Minimum and Maximum Length Scales in Cosmology}  

\author{Seyen Kouwn}
\email{seyen@skku.edu}
\affiliation{
 Department of Physics,~BK21 Physics Research Division,\\
~Institute of Basic Science,\\
Sungkyunkwan University, Suwon 440-746, Korea
}

\date{\today}


\begin{abstract}
We investigate the cosmological implications of the generalized and extended uncertainty principle (GEUP), and 
whether it could provide an explanation for the dark energy. 
The consequence of the GEUP is the existence of a minimum and a maximum length,
which can in turn modify the entropy area law and also modify the Friedmann equation.
The cosmological consequences are studied 
by paying particular attention to the role of these lengths. 
We find that the theory allows a cosmological evolution 
where the radiation- and matter-dominated epochs are 
followed by a long period of virtually constant dark energy, 
that closely mimics the $\Lambda$CDM model. 
The main cause of the current acceleration arises from 
the maximum length scale $\beta$, governed by the relation $\Lambda\sim -\beta^{-1}W(-\beta^{-1})$. 
Using recent observational data (the Hubble parameters, type Ia supernovae, and baryon acoustic oscillations, 
together with the Planck or WMAP 9-year data 
of the cosmic microwave background radiation), we estimate constraints to
the minimum length scale $\alpha \lesssim 10^{81}$ and the maximum length scale $\beta \sim -10^{-2}$.
\end{abstract}


\maketitle


\section{Introduction}

The observation that the universe is accelerating \cite{Riess:1998cb,Perlmutter:1998np} 
has generated extensive investigations aiming to establish its theoretical foundation.
A promising possible explanation involves invoking the cosmological constant $\Lambda$, which is related to the 
vacuum energy density.
For consistency with existing observations, 
$\Lambda$ must be very small, on the scale of $\sim 10^{-120}$, orders of magnitude smaller than the Planck scale $M_{\rm p}$. However, the exact value gives rise to the ``cosmological constant problem''~\cite{Weinberg:1988cp}.
Another possibility is 
a dynamic dark energy model~\cite{Peebles:1987ek,Sahni:1999gb,Copeland:2006wr}, in which the cosmological constant varies dynamically.
Arguably, 
observational data favor dynamical dark energy models 
over the standard $\Lambda$CDM model~\cite{Zhao:2017cud,Mehrabi:2018oke}.
One possible way to describe dynamic dark energy models is as a
generalization of Heisenberg’s uncertainty principle.
 
Early applications of this principle concerned mainly 
black-hole thermodynamics~\cite{Setare:2004sr,Custodio:2003jp,Kouwn:2009qb}, 
but more recently also cosmological topics,
such as inflation~\cite{Tawfik:2014dza,Mohammadi:2015upa},
non-singular universe construction~\cite{Salah:2016kre,Khodadi:2016gyw} and 
the dark energy model~\cite{Majumder:2013fza,Jalalzadeh:2013zwa}.
The underlying idea is that
a generalization of the principle can modify the entropy-area relation in thermodynamics, 
thereby introducing corrections to the cosmological evolution equation.
One well-known example is the ``generalized uncertainty principle''~(GUP)~\cite{Kempf:1994su},
 $\Delta x \Delta p \geq 
1/2 +  \alpha l_p^2/2 \Delta p^2$,
which allows the introduction of quantum-gravity
into ordinary quantum mechanics via the deformation of the Heisenberg uncertainty principle.
Such a deformation implies the existence of a minimum length, 
$\Delta x_{\rm min} \sim \alpha^{1/2} l_p$,
and is expected to have been most apparent in the early universe or in the high-energy regime.
Another possible generalization is 
the ``extended uncertainty principle''~(EUP)~\cite{Bambi:2007ty},
$\Delta x \Delta p \geq 1/2 + \beta/L^2_{\rm x}\Delta x^2$,
where $\beta$ is a dimensionless parameter and $L_{\rm x}$ is an unknown fundamental length scale.
In contrast to the GUP, the EUP implies the existence of a minimum momentum with positive values of $\beta$, 
$\Delta p_{\rm min} \sim \beta^{1/2}/L_{\rm x}$,
and is predicted to be most apparent at later times in the universe.
However, as mentioned in~\cite{Bolen:2004sq}, 
it is interesting that
a positive cosmological constant can only result from 
negative values of $\beta$, thus
contradicting the EUP prediction of a minimum momentum.
In such a case, a position measurement may not exceed 
an unknown length scale, i.e., the maximum length 
$\Delta x_{\rm max} \sim L_{\rm x}/\beta^{1/2}$.
By combining the EUP and GUP (GEUP) we obtain a more general form~\cite{Kempf:1994su,Bojowald:2011jd}
\begin{align}\label{GEUPun}
\Delta x \Delta p \geq 
{1 \over 2} 
\left(1 +  \alpha l_p^2 \Delta p^2 + { \beta \over L_{\rm x}^2}\Delta x^2  \right)
\,.
\end{align}
This formulation of the GEUP~\eqref{GEUPun} predicts the existence of 
both a minimum and a maximum length (with negative values of $\beta$).
It is worth mentioning that 
these modified Heisenberg uncertainty principles (i.e., the GUP, EUP, or GEUP),
can yield a correction to the Bekenstein--Hawking entropy of a black hole~\cite{Medved:2004yu,Majumder:2011xg}.

The connection between thermodynamics and gravity 
was first investigated by Bardden, Carter, and Hawking~\cite{Bardeen:1973gs}.
There has since then been an abundant literature on, e.g., the Rindler space-time~\cite{Jacobson:1995ab} and the Friedmann--Robertson--Walker~(FRW) universe~\cite{Akbar:2006kj}.
With regard to the Rindler space-time, Jacobson found that 
the Einstein equation can be derived from
the thermodynamic relation between heat, entropy, and temperature: $dQ=TdS$,
where $dQ$ is the energy flux and $T$ is the Unruh temperature,
which are detected by an accelerated observer located
just within the local Rindler causal horizons.
The FRW universe, on the other hand, 
assumes that the apparent horizon $\tilde{r}_{A}$ has an associated entropy $S=A/4G$
and a temperature $T=\kappa / 2\pi$ in Einstein gravity,
where $A$ and $\kappa$ are, respectively, the area and surface gravity of the apparent horizon.
Akbar and Cai derived the differential form of the Friedmann equation for a FRW universe
from the first law of thermodynamics at the apparent horizon, i.e., $dE=TdS+WdV$, 
where $E$ is the total energy density of matter existing within the apparent horizon,
$V$ is the volume contained within the apparent horizon,
and the work density $W=(\rho-p)/2$ is a function of the energy density $\rho$
and the pressure $p$ of matter in the universe.
A modified Friedmann equation was recently suggested~\cite{Awad:2014bta} based on a corrected entropy formula
that is potentially useful within the context of cosmology.

The purpose of the present study is to consider cosmology  
within the framework of the GEUP, i.e., by considering the minimum and maximum lengths,
and to compare the results with observations.
The GEUP involves two parameters that can be constrained by measurements.

This paper is organized as follows: 
In Section~\ref{sec2}, 
we investigate the influence of the GEUP on thermodynamics,
and obtain a corrected Friedmann equation for the FRW universe.
In Section~\ref{sec3}, 
we investigate the effects of the GEUP length-scale parameters $\alpha$ and $\beta$,
and find that the theory is consistent with the long acceleration phase currently undergone by the universe.
Section~\ref{sec4} presents 
observational constraints on our model parameters. 
Section~\ref{sec5} closes with discussions and concluding remarks.

\section{Modified Friedmann Equations}\label{sec2}
This section
presents calculations of the modified Friedmann equations
within the framework of the GEUP to describe cosmological effects.
The outcome of the GEUP~\eqref{GEUPun} is the modified momentum uncertainty
\begin{align}
\Delta p 
&\geq {\Delta x \over \alpha l_p^2}
\left(
1-\sqrt{ 1-{l_p^2 \over L_{\rm x}^2}\alpha \beta - {\alpha l_p^2 \over \Delta x^2}  }
\right) \,, \\
&\simeq 
{ 1 \over 2 \Delta x}\left(
1+{ \alpha l_p^2 \over 4 \Delta x^2 } + {\beta \over L_{\rm x}^2}\Delta x^2
\right) \,,
\end{align}
with the Taylor expansion calculated at $\alpha=\beta=0$.
As noted in~\cite{Medved:2004yu}, the Heisenberg uncertainty principle $\Delta p > 1/\Delta x$ 
can be rewritten in terms of a lower bound to the energy ($E>1/\Delta x$),
which in the case of the GEUP becomes 
\begin{align}\label{}
E \geq {1\over 2 \Delta x}\left(
1+{ \alpha l_p^2 \over 4 \Delta x^2 } + {\beta \over L_{\rm x}^2}\Delta x^2
\right)
\,.
\end{align}
When a black hole absorbs or emits
a classical particle of energy $E$ and size $R$,
the minimal change in the surface area of the black hole is $\Delta A_{\rm min} \geq 8\pi l_p E R$.
Arguably, the size of a quantum particle
cannot be smaller than $\Delta x$ ~\cite{AmelinoCamelia:2005ik}, 
which would imply the existence of a finite bound $\Delta A_{\rm min} \geq 8\pi l_p E \Delta x$.
Thus, considering the GEUP, we obtain
\begin{align}\label{Amin}
\Delta A_{\rm min} \geq 4\pi l_p \left( 
1+{ \alpha l_p^2 \over 4 \Delta x^2 } + {\beta \over L_{\rm x}^2}\Delta x^2
\right)
\,.
\end{align}
$\Delta x$ is the position uncertainty of a photon 
which can be associated with the black hole radius, $\Delta x = 2 r_s$, 
where $r_s$ is the Schwarzschild radius.
Given the surface area of the black hole, $A=4\pi r_s^2$, the relation between $A$ and $\Delta x$
can be expressed as $\Delta x^2 = A/\pi$.
Substituting this equation into \eqref{Amin}, 
the minimal area change becomes
\begin{align}\label{}
\Delta A_{\rm min} \geq 4\pi l_p \lambda \left( 
1+{ \pi \alpha l_p^2 \over 4 A } + {\beta \over \pi L_{\rm x}^2} A
\right)
\,,
\end{align}
where $\lambda$ is the calibration factor that is determined from the Bekenstein--Hawking entropy formula.
The entropy of the black hole is assumed to 
depend on its surface area. Also, given that the entropy increases by a factor of $\ln 2$ at least,
regardless of the value of the area, we have
\begin{align}\label{dsda}
{dS \over dA}
={\Delta S_{\rm min} \over \Delta A_{\rm min}} 
= {1\over 4 l_p^2 } \left( 
1+{ \pi \alpha l_p^2 \over 4 A } + {\beta \over \pi L_{\rm x}^2} A
\right)^{-1}
\,,
\end{align}
where $\ln 2 / \lambda = \pi$, as mentioned above.
Integrating \eqref{dsda}, the GEUP-corrected entropy is
\begin{align}\label{mds}
S
= {A\over 4 l_p^2 } \left( 
1-{ \pi \alpha l_p^2 \over 4 A }\ln\left({A\over 4 l_p^2}\right) 
- {\beta \over 2\pi L_{\rm x}^2} A
\right)
\,.
\end{align}
We note that the modified Bekenstein--Hawking entropy~\eqref{mds}
arises from the existence of the minimum and maximum lengths.

Based on the ``apparent horizon'' approach~\cite{Akbar:2006kj}, 
we derived the modified Friedmann equations
with the modified entropy~\eqref{mds} applied to the first law of thermodynamics, $dE=TdS+WdV$.
Thus, we considered that space-time geometry is characterized by the FRW metric
\begin{align}\label{frw}
ds^2 = -dt^2 + a^2\left(
{dr^2 \over 1- kr^2}+r^2 d\Omega_2^2 
\right) \,,
\end{align}
where $a$ is a scaling factor of our universe, 
and the values of the spatial curvature constant $k=+1$, $0$, or $-1$ correspond, respectively, to a closed, flat, or open universe.
Using spherical symmetry, the metric~\eqref{frw} can be rewritten as 
\begin{align}
ds^2 = h_{ab}dx^a dx^b + \tilde{r}^2 d\Omega_2^2 \,,
\end{align}
where $x^0=t$, $x^1=r$ and $\tilde{r} = ar$,
and the two-dimensional metric $h_{ab}= {\rm diag}(-1,a^2/(1-kr^2))$.
In the FRW universe, a dynamic horizon always exists because it is a local quantity of space-time,
which is a marginally trapped surface with vanishing expansion.
It is determined by the relation $h^{ab}\partial_a \tilde{r}\partial_b \tilde{r}$, 
which yields the radius of the apparent horizon
\begin{align}
\tilde{r}_{\rm A}^2 = {1\over H^2 + k/a^2} \,,
\end{align}
where $H = \dot{a}/a $ is the Hubble parameter.
By assuming that matter in the FRW universe forms a perfect fluid 
with four-velocity $u^\mu$, the energy-momentum tensor can be written 
\begin{align}
T_{\mu\nu} = (\rho+p)u_\mu u_\nu + p g_{\mu\nu}
 \,,
\end{align}
where $\rho$ is the energy density of the perfect fluid and $p$ is its pressure.
The energy conservation law, $\nabla_{\mu}T^{\mu\nu}=0$, yields the continuity equation
\begin{align}
\dot{\rho} + 3H(\rho+p)=0
 \,.
\end{align}
According to the main results of \cite{Akbar:2006kj,Awad:2014bta}, 
applying the first law of thermodynamics to the apparent horizon of the FRW universe yields the corresponding Friedmann equations
\begin{align}
{8\pi G \over 3} \rho &= -16 \pi G\int {S'(A) \over A^2 } dA \,, \label{mdfrw1} \\
-\pi (\rho+p) &=  S'(A) \left( \dot{H} - {k \over a^2} \right) \label{mdfrw2} \,,
\end{align}
where $A$ is the area of the apparent horizon, given by
\begin{align}
A = 4\pi \tilde{r}_{\rm A}^2 = { 4\pi \over H^2 + {k\over a^2} }
\,.
\end{align}
Substituting the modified entropy \eqref{mds} into the modified Friedmann equations \eqref{mdfrw1} and \eqref{mdfrw2},
we obtain
\begin{align}
H^2 + {k \over a^2} &= {1\over 3 M_p^2} \left( 
\rho  + \rho_{\rm x} \right) \,, \label{mdfrw3} \\
\dot{H} - {k \over a^2} &= -{1\over 2M_p^2}
\left(
\rho+p
+ \rho_{\rm x} + p_{\rm x} 
\right) \,, \label{mdfrw4}
\end{align}
where $\rho_{\rm x}$ and $p_{\rm x}$ are, respectively, the energy density and pressure originating from the
GEUP. We interpret $\rho_{\rm x}$ and $p_{\rm x}$ as the ``dark energy density'' and the ``dark pressure'', respectively:
\begin{align}
\rho_{\rm x} 
&= 
\tilde{\alpha} \left(H^2+{k\over a^2} \right)^{2} 
+ M_p^4\tilde{\beta} \ln \left(  { H^2+{k\over a^2} \over 8 \pi^2 M_p^2 } \right) \,, \label{rhox}\\
\rho_{\rm x} + p_{\rm x} 
&= {-2 \over 3}\left(
2\tilde{\alpha} \left(H^2+{k\over a^2} \right)
+{M_p^4 \tilde{\beta} \over H^2+{k\over a^2}  }
\right)
\left( \dot{H} - {k \over a^2} \right) \,,\label{rhopx}
\end{align}
where $\tilde{\alpha} \equiv {3 \alpha \over 256\pi} $ and 
$\tilde{\beta} \equiv { 12 \beta \over M_p^2 L_{\rm x}^2 } $ for conciseness, and
$\rho_{\rm x}$ and $p_{\rm x}$ satisfy the energy conservation law
\begin{align}
\dot{\rho}_{\rm x} + 3H\left( \rho_{\rm x} + p_{\rm x} \right) = 0\,.
\end{align}
Note that setting $\beta=0$ in \eqref{GEUPun} amounts to the GUP model,
with the corresponding dark energy density $\sim \tilde{\alpha}H^4$.
In this case, the energy density, which scales like $H^4$,
cannot explain the acceleration of the present universe.
Even if the energy density were proportional to $H^2$, 
it would not account for the present acceleration either~\cite{Maggiore:2010wr,Basilakos:2009wi}.
This is due to the energy density decreasing very quickly.
GUP alone can therefore not explain the current accelerating universe.
However, by further imposing a maximum length,
the energy density acquires a logarithmic term $\sim \ln H$.
Conceptually, 
the exponent of $\ln H$ is nearly zero,
so that the change in $\ln H$ is not large (i.e., $\ln H \sim$ constant).
It is therefore possible to explain the acceleration of the universe
via the $\ln H$ term, derived from the maximum length. Indeed, observations suggest that 
the dark energy density is 
almost constant at present.

\section{Cosmology}\label{sec3}
This section analyzes the evolution equations by assuming that the universe, at each stage of its existence, is dominated by a barotropic perfect fluid with a constant equation-of-state
parameter $w_i = \rho_i/p_i(i=m,r)$ and, later, by $\rho_{\rm x}$.
The evolution equation is then 
\begin{align}\label{eqH2full}
H^2 + {k \over a^2} &= {1\over 3 M_p^2} \left( 
\rho_{r} + \rho_{m}  + \rho_{\rm x} \right) \,.
\end{align}

\begin{figure*}[ht]
\begin{center}
\resizebox{\textwidth}{!}{
\includegraphics{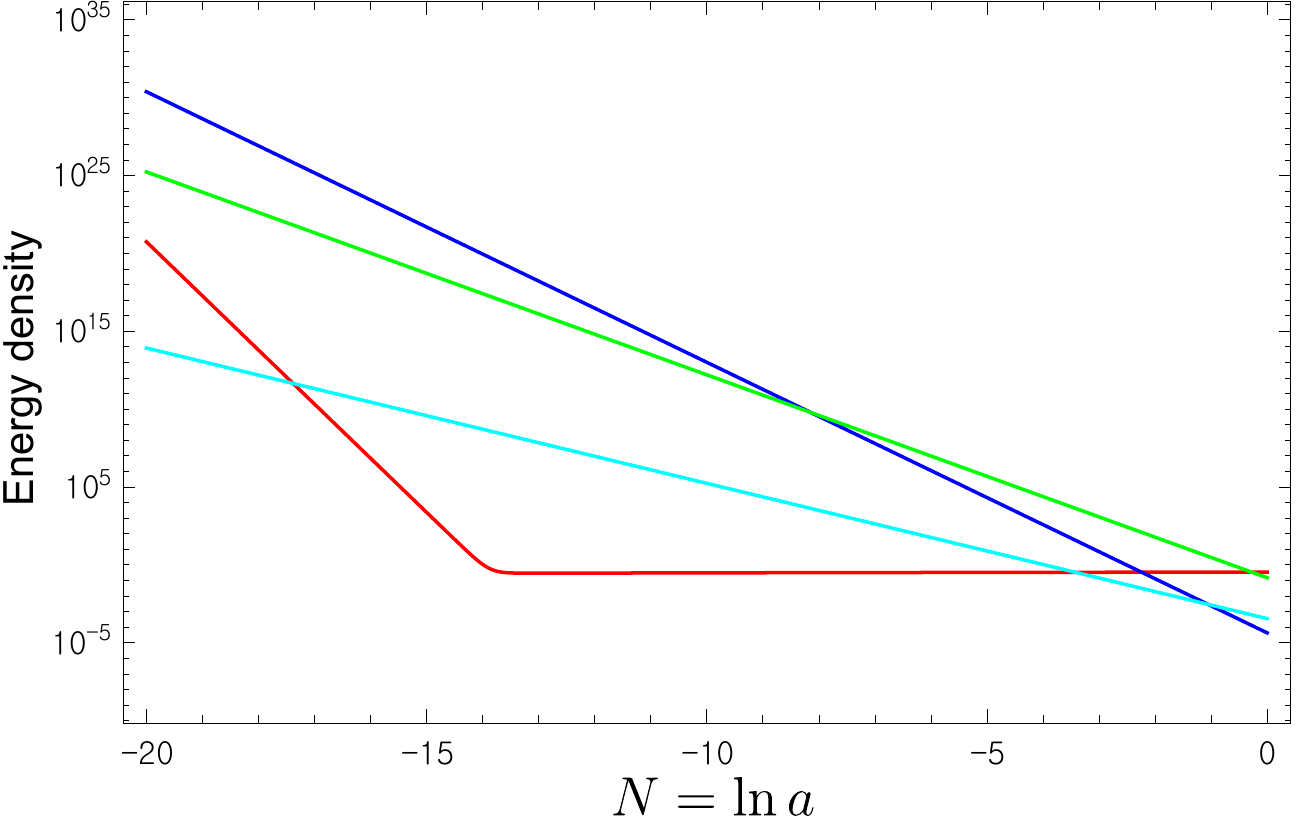} \quad \includegraphics{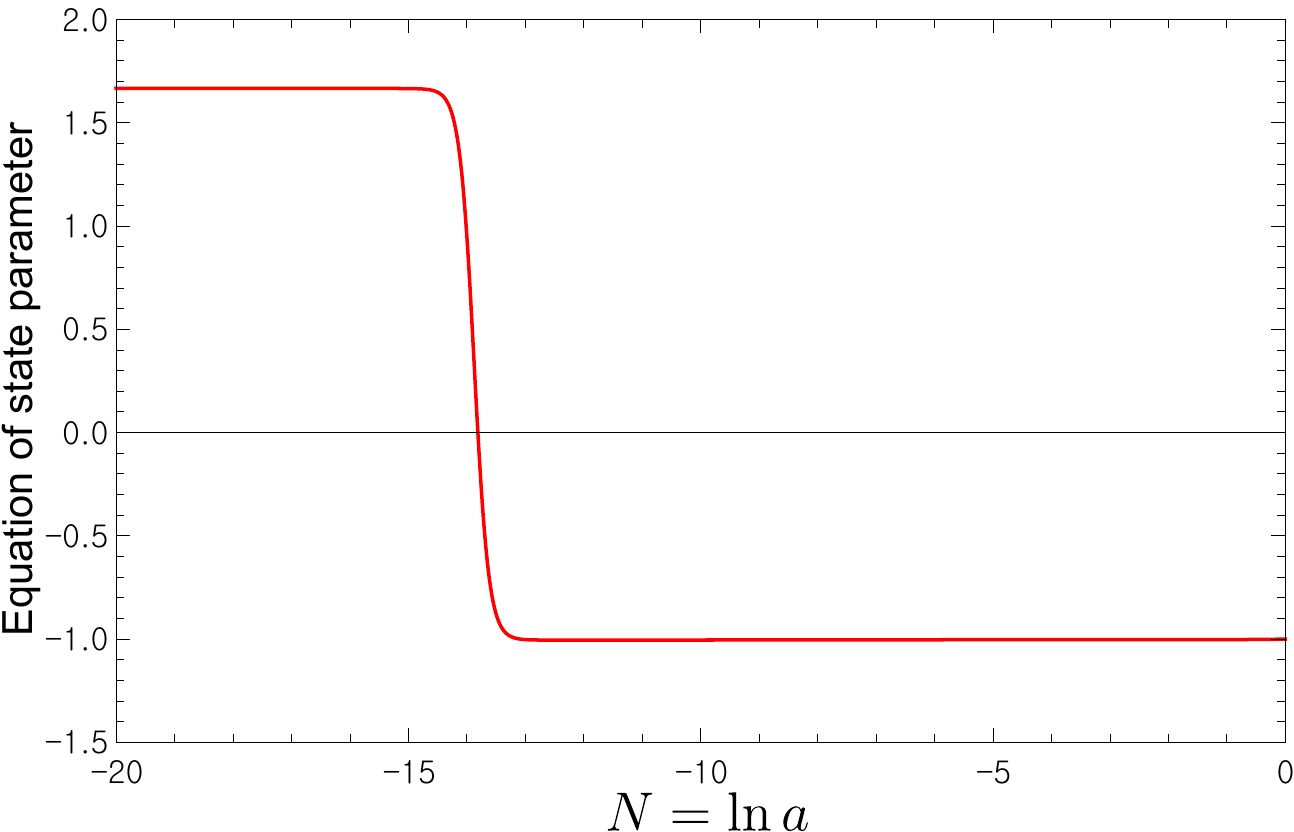}
}
\end{center}
\caption{\small
(Left) Evolution of the energy density of the GEUP(red), radiation(blue), matter(green) and curvature(cyan). 
(Right) Evolution of the equation-of-state parameter for the GEUP. 
In both panels, we have used
$\log_{10}\hat{\alpha}=-40$, $\hat{\beta}=-0.0012$ $\Omega_m h^2 = 0.144$ and $\Omega_k h^2 = -0.0003$.
}\label{fig:energyeos}
\end{figure*}

\subsection{Early-time approximation}
It is assumed that, in the early stages of the universe, 
the energy densities of matter and of the dark energy were 
negligibly small compared to that of radiation:
\begin{align}\label{radcon}
\rho_{\rm x} \ll \rho_{r}\,,\quad
\rho_{\rm m} \ll \rho_{r}\,.
\end{align}
The Hubble parameter $H$ is hence given by
\begin{align}
3 M_p^2 H^2 \simeq {\rho_{r,0} \over a^4}\,,
\end{align}
where $\rho_{r,0}$ is the present value of the radiation energy density. 
At these early stages, 
the first term dominated in \eqref{rhox}, i.e., $\rho_{\rm x} \sim \tilde{\alpha} H^4$,
so that the conditions \eqref{radcon} yield the constraint
\begin{align}\label{alphacond}
\tilde{\alpha} \rho_{r} \ll 1
\,.
\end{align}
Note that
big-bang nucleosynthesis (BBN) imposes an upper bound 
on any source of additional energy density present 
in the universe at the time of BBN ~\cite{Maggiore:2010wr,Iocco:2008va},
giving the upper limit for that source $\rho_{\rm x}$ in the GEUP model
\begin{align}\label{bbncond}
\left( {\rho_{\rm x} \over \rho_r }\right)_{\rm BBN} \lesssim {\cal O}(0.1) 
\,,
\end{align}
where the BBN epoch is $a \sim 10^{-10}$.
Thus, the conditions~\eqref{radcon} and~\eqref{alphacond} are naturally satisfied by the BBN constraint.
According to \eqref{rhox} and \eqref{rhopx}, 
the dark energy and pressure are given by
\begin{align}\label{earlyrhox}
\rho_{\rm x} \sim {\tilde{\alpha} \over 9} \rho_r^2 \,, \quad
p_{\rm x} \sim {5\tilde{\alpha} \over 27} \rho_r^2\,.
\end{align}
We calculate the equation-of-state parameter
during the radiation-dominated epoch as
\begin{align}
w_{\rm x} \equiv {p_{\rm x} \over \rho_{\rm x} } \simeq {5\over 3}
\,.
\end{align}

\subsection{Late-time approximation}
This subsection considers the later stages of the universe,
which are dominated by the dark energy:
\begin{align}\label{darkcon}
\rho_{r} \ll \rho_{\rm x}\,,\quad
\rho_{m} \ll \rho_{\rm x}\,,
\end{align}
The Hubble parameter $H$ is given by
\begin{align}
3 M_p^2 H^2 \simeq \rho_{\rm x} \,.
\end{align}
In this case, 
the second term is dominant in \eqref{rhox}, $\rho_{\rm x} \sim \tilde{\beta} \ln H$,
because the first term $\sim H^4$ decreases rapidly.
The dark energy density \eqref{rhox} can then be rewritten
\begin{align}\label{rhoxeq}
\rho_{\rm x} \sim \tilde{\beta} \ln \rho_{\rm x} \,.
\end{align}
In order to satisfy \eqref{rhoxeq}, 
the dark energy density should be almost constant
for a given constant value of $\tilde{\beta}$
during this dark energy dominated epoch.
Its value can be obtained as
\begin{align}\label{rhoxdarkdomi}
\rho_{\rm x} \sim -\tilde{\beta}W\left(-\tilde{\beta}^{-1}\right)
 \,,
\end{align}
where $W$ is the Lambert function,
defined as the solution to the equation
$W(x)e^{W(x)}=x$. 
We note that, according to \eqref{rhoxdarkdomi}, 
the dark energy density becomes essentially static
during this dark energy dominated epoch.
This makes the corresponding pressure approximately $p_{\rm x} \simeq -\rho_{\rm x}$
because $\dot{\rho}_{\rm x} \simeq 0$.
In addition, we consider only negative values of $\tilde{\beta}$, as estimated from observations,
in order that \eqref{rhoxdarkdomi} yield a positive dark energy density.

\begin{figure}[ht]
\begin{center}
\resizebox{\columnwidth}{!}{ 
\includegraphics{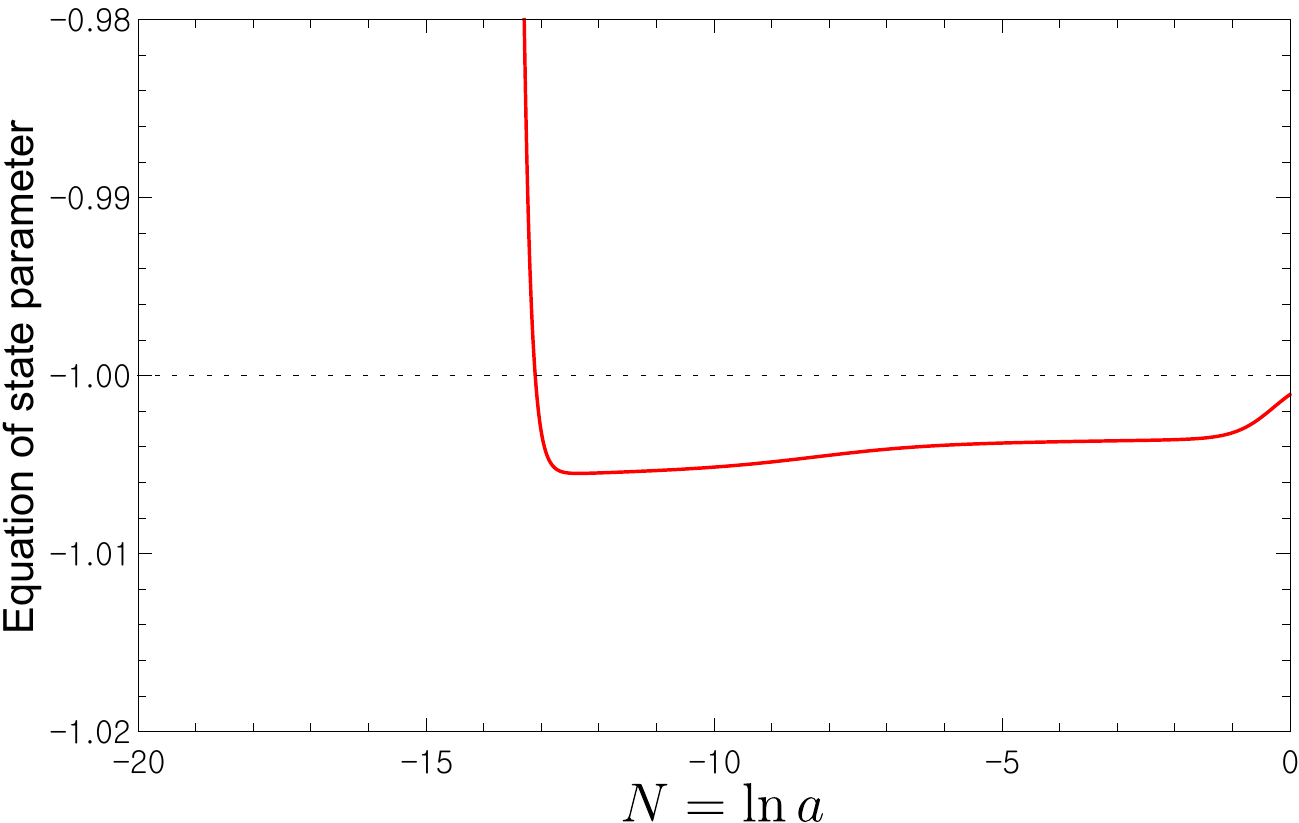} 
}
\end{center}
\caption{\small
Evolution of the equation-of-state parameter near $w_{\rm x}=-1$.
The same model parameters as in Fig.~\ref{fig:energyeos} were used.
}
\label{fig:eos2}
\end{figure}

Some comments are in order at this point. 
Firstly, one can notice that the dark energy density in each epoch, 
resulting from the GEUP, mostly depends on 
 the minimum length scale $\tilde{\alpha}$ or the maximum length scale $\tilde{\beta}$,
according to \eqref{earlyrhox} and \eqref{rhoxdarkdomi}, respectively.
These results are consistent with a previous study~\cite{Zhu:2008cg}.
Secondly, Fig.~\ref{fig:energyeos} shows that 
the dark energy density decreases as $a^{-8}$ during the early epoch,
and remains almost constant subsequently.
Thus, our numerical results are consistent with
approximate analytical solutions.
Finally, a notable feature is that 
the equation-of-state parameter $w_{\rm x}$ dynamically crosses over the value $-1$ (phantom crossing), with
its values being slightly negative at present (Fig.~\ref{fig:eos2}),
consistent with observations~\cite{Zhao:2017cud,Mehrabi:2018oke}.

\section{Observational Constraints}\label{sec4}
This section 
discusses the parameter estimation for our model, by Markov Chain Monte Carlo(MCMC) simulations and 
using the most recently available cosmological data,
to investigate whether or not
it can be distinguished from the $\Lambda$CDM model.
For this purpose,
we used the recent observational data, e.g., from type Ia supernovae (SN),
baryon acoustic oscillations (BAO) imprinted in the large-scale structure of galaxies,
cosmic microwave background radiation (CMB), and Hubble parameters [$H(z)$].
The likelihood distributions for the model parameters were
derived using the maximum-likelihood method.
This method involves exploring the parameter space covered by the vector $\mbox{\boldmath $\theta$}$ in random directions. The choice of parameters that is favored by the observational data is determined by deciding whether to accept or reject a given randomly chosen parameter vector in successive iterations. This decision is made using
the probability function
$P(\mbox{\boldmath $\theta$}|\mathbf{D}) \propto \exp(-\chi^2/2)$,
where $\mathbf{D}$ denotes the data 
and $\chi^2= \chi_{H(z)}^2+\chi_\textrm{SN}^2 + \chi_\textrm{BAO}^2 + \chi_\textrm{CMB}^2
$ is the sum of individual chi-squares
for the $H(z)$, SN, BAO, and CMB data.
Full details are provided in Section 4 of~\cite{Kouwn:2015cdw}.
For numerical analysis,
it is convenient to rewrite the evolution equation \eqref{eqH2full}
in terms of
$N\equiv \ln a$ as follows:
\begin{align}
\hat{H}^2  &= \Omega_r h^2 e^{-4N} + \Omega_m h^2 e^{-3N}
+ \Omega_{\rm x}h^2(N) + \Omega_k h^2 e^{-2N} \,,
\end{align}
and 
\begin{align}
\Omega_{\rm x}h^2(N) 
&=
\hat{\alpha} \left( \hat{H}^2-{\Omega_kh^2 e^{-2N}}\right)^{2} \nn \\
&\quad\quad
+\hat{\beta}  \ln \left( 
{H_0^2 \left(\hat{H}^2-{\Omega_kh^2 e^{-2N}} \right) \over 4\pi M_p^2 h^2 }
\right)
\,,
\end{align}
where we have introduced the following dimensionless quantities:
\begin{align}\label{eqdimless}
\hat{H}^2 &\equiv {H^2 h^2 \over H_0^2}\,, \quad
\Omega_r \equiv { \rho_{r,0} \over 3 M_p^2 H_0^2}\,, \quad
\Omega_m \equiv { \rho_{m,0} \over 3 M_p^2 H_0^2}\,, \nn \\
\Omega_k &\equiv { -k \over 3 M_p^2 H_0^2}\,, \quad
\hat{\alpha} \equiv {\tilde{\alpha} H_0^2 \over 3 M_p^2 h^2}\,, \quad
\hat{\beta} \equiv {\tilde{\beta} M_p^2 h^2 \over 3 H_0^2} \,.
\end{align}
$H_0$ is the present value of the Hubble parameter,
usually expressed as $H_0=100 \,h\, {\rm km}\,\,{\rm s}^{-1}{\rm Mpc}^{-1}$, and
$\Omega_r$ and $\Omega_m$ are, respectively, the radiation- and matter-density parameters in the present universe.
For the radiation density, 
we used $\Omega_r h^2 = 4.17587\times 10^{-5}$ for WMAP9 and $\Omega_r h^2 = 4.17893\times 10^{-5}$ for PLANCK~\cite{Kouwn:2015cdw}.
Notice that the background dynamics is completely determined by the set of parameters
$(\log_{10}\hat{\alpha}, \hat{\beta}, \Omega_m, \Omega_k)$.
We also need the baryon density parameter ($\Omega_b$) to compare our model with the BAO and CMB data, 
giving five free parameters in total:
$\mbox{\boldmath $\theta$}
=(\log_{10}\hat{\alpha}, \hat{\beta}, \Omega_b h^2, \Omega_m h^2, \Omega_k h^2)$.
It should be noted that the Hubble constant ($H_0$) is no longer a free parameter
because it can be derived from the equations starting from a given set of chosen parameters.
We take the priors for the free parameters as follows:
$\log_{10}\hat{\alpha} = [-50, -37]$, $\hat{\beta} = [-0.0020,-0.0010]$,
$\Omega_b h^2 = [0.015, 0.030]$, $\Omega_m h^2 = [0.11, 0.15]$ and $\Omega_k h^2 = [-0.1, 0.1]$.

\begin{table*}[ht]
\begin{center}
\caption{Summary of parameter constraints and derived parameters.
The confidence levels are 68\% unless otherwise stated.}
\label{table:result1}
\begin{tabular*}{\textwidth}{@{\extracolsep{\fill}}c||cc|cc}
\hline
\hline
   &  \multicolumn{2}{c}{GEUP Model}  &  \multicolumn{2}{c}{$\Lambda{\rm CDM}$ Model} \\
\hline
 & $H(z)$ + SN + BAO  &  $H(z)$ + SN + BAO & $H(z)$ + SN + BAO & $H(z)$ + SN + BAO \\
 & + WMAP9            & + PLANCK           & +WMAP9            & +PLANCK\\

\hline

$~H_0 \quad$ & $68.86^{+0.88}_{-0.88}$ & $69.11^{+0.86}_{-0.86}$
      & $68.87^{+0.94}_{-0.94}$ & $69.08^{+0.83}_{-0.82}$\\

$~\Omega_m h^2 \quad$ & $0.1360^{+0.0033}_{-0.0034}$ & $0.1440^{+0.0022}_{-0.0022}$
               & $0.1359^{+0.0033}_{-0.0034}$ & $0.1438^{+0.0022}_{-0.0024}$ \\

$~\Omega_b h^2 \quad$  & $0.02451^{+0.00051}_{-0.00056}$ & $0.02396^{+0.00029}_{-0.00031}$
                & $0.02453^{+0.00054}_{-0.00054}$ & $0.02397^{+0.00030}_{-0.00031}$\\

$~\Omega_k \quad$  & $-0.0077^{+0.0039}_{-0.0038}$ & $-0.0011^{+0.0028}_{-0.0028}$
             & $-0.0077^{+0.0038}_{-0.0038}$ & $-0.0012^{+0.0028}_{-0.0028}$\\

$\log \hat{\alpha} \quad$  & $< -37~({\rm BBN})$  & $< -37~({\rm BBN})$
           & - & - \\

$~\hat{\beta} \quad$ & $-0.00122^{+0.00004}_{-0.00004}$  & $-0.00119^{+0.00004}_{-0.00004}$
    & -                        & -\\

$~\Omega_\Lambda h^2 \quad$ & -                            & -
                     & $0.342^{+0.012}_{-0.012}$ & $0.334^{+0.011}_{-0.011}$ \\

\hline
$~\chi^{2}_\textrm{min} \quad$  & 584.278 & 590.312
                                & 584.344 & 590.502 \\
$~\chi^{2}_\nu \quad$  & 0.94850 & 0.95830
                       & 0.94861 & 0.95861 \\
\hline
\hline
\end{tabular*}
\end{center}
\end{table*}

\begin{figure}[ht]
\begin{center}
\resizebox{\columnwidth}{!}{ 
\includegraphics{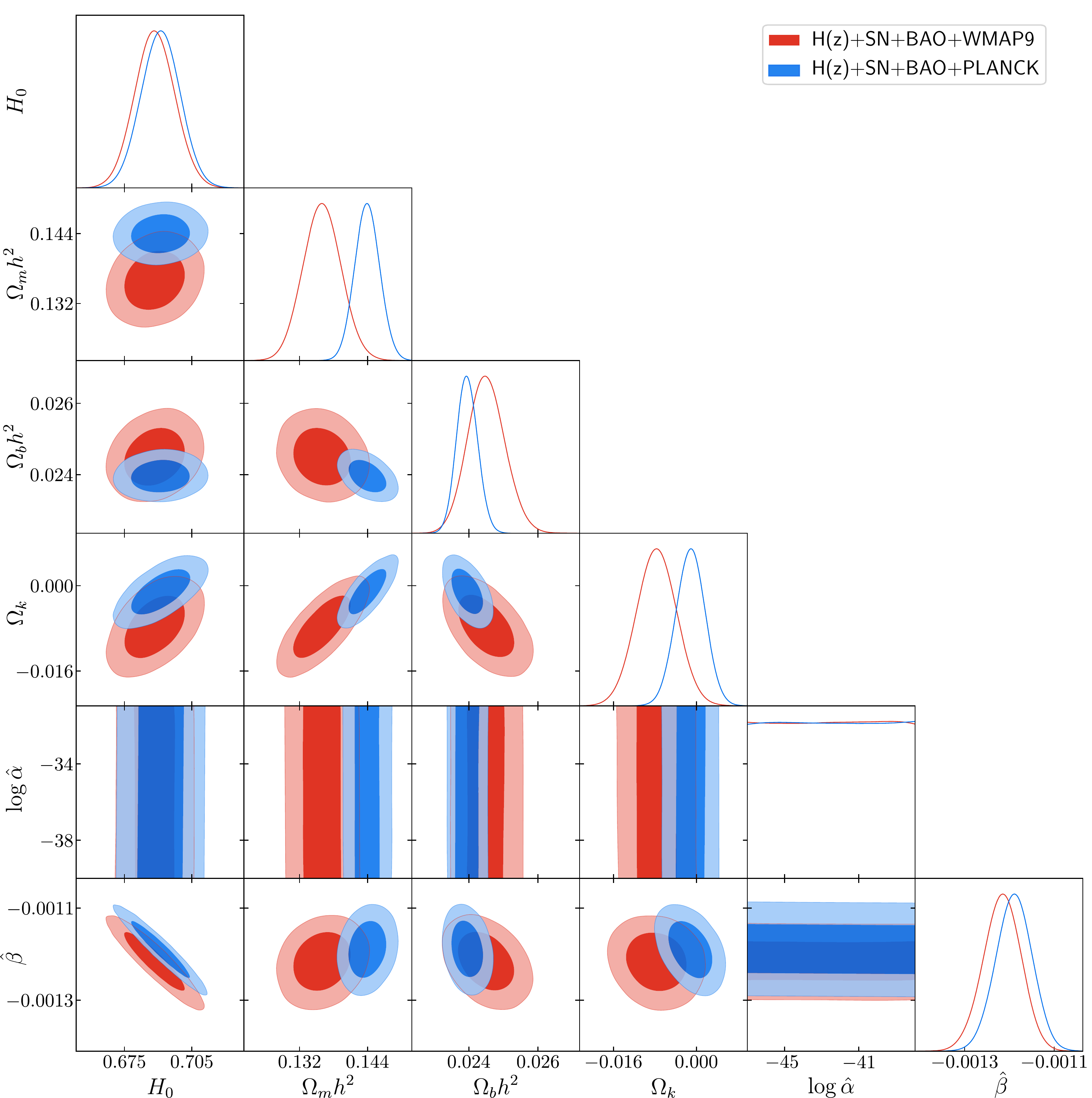} 
}
\end{center}
\caption{\small
Marginalized likelihood distributions of parameters for different dataset combinations. 
WMAP9 and PLANCK refer to H(z) + SN+BAO+WMAP9 and H(z) + SN+BAO+PLANCK, respectively. 
The contours indicate the 68\% and 95\% confidence limits.
}
\label{fig:cont}
\end{figure}

\subsection{Results}

We explored the allowed parameter ranges of our dark energy model by using the recent observational data within the MCMC method. In the calculation, 
we used $\log \hat{\alpha}$, $\hat{\beta}$, $\Omega_m h^2$, $\Omega_b h^2$ and $\Omega_k h^2$ 
as free parameters. 
Table~\ref{table:result1}
summarizes the parameter mean values and $68\%$ confidence limits,
and Fig.~\ref{fig:cont} shows the marginalized likelihood distributions of the parameters.
We can see that the result obtained with the Planck data
show slightly tighter constraints on the model parameters.
Parameter $\alpha$, within the BBN constraint~\eqref{bbncond} ($\log \hat{\alpha} < -37$),
is uniformly distributed because it is uncorrelated with any of the other model parameters.
In other words, the parameter estimation results are relatively unaffected by the presence of the minimum length scale $\alpha$.
The constraints on $\log \hat{\alpha}$ in Table~\ref{table:result1} are therefore only cited as BBN.
The best-fit locations in the parameter space are
\begin{align}
&(\Omega_m h^2, \Omega_b h^2, \Omega_k , \log_{10} \hat{\alpha}, \hat{\beta}) \nn \\
&\quad =(0.1359,0.02447,-0.0037,-38.3,-0.00121) \,,
\end{align}
with a minimum chi-square of $\chi_\textrm{min}^2 = 584.278$ for the $H(z)$+SN+BAO+WMAP9, 
and
\begin{align}
&(\Omega_m h^2, \Omega_b h^2, \Omega_k , \log_{10} \hat{\alpha}, \hat{\beta}) \nn \\
&\quad =(0.1438,0.02395,-0.0006,-41.4,-0.00119) \,,
\end{align}
with $\chi_\textrm{min}^2 = 590.312$ for $H(z)$+SN+BAO+PLANCK.

To assess the goodness of fit of our model, Table \ref{table:result1}
displays the parameter constraints for the $\Lambda\textrm{CDM}$ model
and lists, for each case, the value of the minimum reduced chi-square ($\chi_\nu^2$). This is defined as $\chi_\nu^2=\chi_{\textrm{min}}^2 /\nu$,
where $\nu=N-n-1$ is the number of degrees of freedom and $N$ and $n$ are the numbers of data points
and free model parameters, respectively.
In our analysis, $N=621$, and $n=5$ for our model
and $n=4$ for the $\Lambda\textrm{CDM}$ model.
Our model fits the observational data slightly better, 
with smaller values of $\chi_\textrm{min}^2$ and $\chi_\nu^2$.
We note that, for our model to be compatible with observations,
$\log_{10} \hat{\alpha}$ must be smaller than $\sim -37$ (BBN constraint)
and $\hat{\beta}$ should be close to $\sim -10^{-3}$.
Identifying the unknown maximum length scale $L_{\rm x}$ with 
the Hubble radius $1/H_0$ of the current universe,
\eqref{eqdimless} gives
\begin{align}
\alpha \lesssim 10^{81}\,, \quad
\beta = \hat{\beta}/2 \sim -10^{-2}
\,.
\end{align}
The BBN~\eqref{bbncond} does not impose a strong constraint
on the minimum length scale $\alpha$ 
compared with other results
(i.e., $\alpha\lesssim 10^{34}$ from the electroweak length scale~\cite{Das:2008kaa},
$\alpha\lesssim 10^{36}$ from measurement of the Lamb shift~\cite{Brau:1999uv},
and $\alpha\lesssim 10^{50}$ from measurement of Landau levels~\cite{Wildo:1997}).

\section{Conclusion}\label{sec5}

To conclude, 
we investigated the cosmological implications 
of minimum and maximum lengths in the FRW universe, 
giving special attention to the potential significance of these lengths in relation with the dark energy. 
We found that the theory is consistent with the long period of acceleration which the universe is presently undergoing, a phase 
that closely mimics the $\Lambda$CDM model, in which the acceleration of the universe is 
due to the existence of the maximum length scale $\beta$ 
governed by the relation $\Lambda\sim -\beta^{-1}W(-\beta^{-1})$. 
A detailed numerical analysis, comparing various available data, predicts that 
the maximum length, governed by $\beta$, is of the order of $\sim -10^{-2}$,
and the minimal length, governed by $\alpha$, is of the order of $\lesssim 10^{81}$.

Some interesting properties of cosmological evolution within the GEUP framework
arise from the existence of the minimum and maximum lengths.
The minimum length introduces a correction term in the Friedmann equation,
with a corresponding energy density that scales as $\sim \tilde{\alpha} H^4$. This is likely to have played
an important role predominantly in the early universe.
On the other hand, the energy density arising from the maximum length is 
given, in practice, by an intriguing relation $\sim \tilde{\beta}\ln H$, 
which became significant mostly in the later universe.
With regard to dynamics, the $\tilde{\beta}\ln H$ term increases very slowly (as compared to a power-law behavior for $H$)
and eventually becomes dominant in the present epoch.
The minimal length scale $\alpha$ does not affect 
the current acceleration of the universe
as long as the BBN constraint is satisfied.
Even setting $\alpha=0$ can explain the present acceleration of the universe via the existence of the maximum length only,
but its equation-of-state parameter will be always in phantom phase as the universe expands.
However, 
by imposing a minimal length, 
the equation-of-state parameter starts from $w_{\rm x}\simeq 5/3$ and 
cross the phantom divide in the intermediate state 
between the radiation- and matter-dominated epochs (Fig.~\ref{fig:eos2}),
and the equation-of-state parameter at present is 
in the phantom phase,
as allowed by observations.


\begin{acknowledgments}
We thank Seokcheon Lee and Chan-Gyung Park for useful discussions.
This work was supported by the Basic Science Research Program through
the National Research Foundation of Korea (NRF), funded by the Ministry of Education (Grant No. NRF-2017R1D1A1B03032970).
\end{acknowledgments}

\end{document}